\documentclass[aps,amsmath, amssymb, prd, twocolumn, superscriptaddress,floatfix,showpacs,bibliography,10pt]{revtex4-1}
\usepackage{graphicx}  % Required for including images
\graphicspath{{figures/}}

\usepackage{caption}
\usepackage{subcaption}

\usepackage{hyperref}			% Creates bookmarks
\usepackage[all]{hypcap}
\usepackage{sistyle}
\usepackage{enumitem}
\usepackage{color}
\usepackage{cancel}

\usepackage{tikz}
\tikzset{dot/.style={shape=circle,fill=black,scale=0.3}}

\usepackage{mathtools}

\newcommand{\coloneq}{\mathrel{\resizebox{\widthof{$\mathord{=}$}}{\height}{ $\!\!\resizebox{1.2\width}{0.8\height}{\raisebox{0.23ex}{$\mathop{:}$}}\!\!=\!\!$ }}}

\begin{document}

\title{Series expansion of the overlap reduction function for scalar and vector polarizations for gravitational wave search with pulsar timing arrays}

\author{Adrian \surname{Bo\^itier}}
\email[]{boitier@physik.uzh.ch}
\affiliation{Physik-Institut, Universit\"at Z\"urich, Winterthurerstrasse 190, 8057 Z\"urich, Switzerland}

\author{Tanguy \surname{Giroud}}
\email[]{tgiroud@student.ethz.ch}
\affiliation{Departement Physik, ETH Z\"urich, R\"amistrasse 101, 8092 Z\"urich, Switzerland}

\author{Shubhanshu \surname{Tiwari}}
\email[]{stiwari@physik.uzh.ch}
\affiliation{Physik-Institut, Universit\"at Z\"urich, Winterthurerstrasse 190, 8057 Z\"urich, Switzerland}

\author{Philippe \surname{Jetzer}}
\email[]{jetzer@physik.uzh.ch}
\affiliation{Physik-Institut, Universit\"at Z\"urich, Winterthurerstrasse 190, 8057 Z\"urich, Switzerland}

\date{\today} 
\newcommand{\shubh}[1]{\textcolor{red}{Shubhanshu: #1}}
\newcommand{\philippe}[1]{\textcolor{green!50!black}{Philippe: #1}}
\begin{abstract}
In our previous work \cite{PTA2} we calculated the overlap reduction function for the tensor polarization without employing the short wavelength approximation, this was done by obtaining a power series of nested sums which is valid for all gravitational wave frequencies and pulsar distances. In this work we generalize the power-series expansion method to vector and scalar polarizations. We have compared our expression for the breathing and vector modes with previous literature. We present for the first time  analytic expressions for the overlap reduction function of the longitudinal mode for all angles between the pulsar pairs. 

%  Pulsar Timing Array (PTA) is one of the most promising technique to detect low-frequency gravitational waves in the near future. In particular, it could be used to study the gravitational wave background in a similar way to the cosmic microwave background. It also opens a new testing ground for alternative metric theories that include non-Einsteinian gravitational wave (GW) scalar and vectorial polarization modes. A recent work \cite{PTA2} established a new methodology to derive a fully analytical expression for one of the most important quantity used to detect the correlations between pulsar signals in PTA: the overlap reduction function, which characterizes the detector dependence on the pulsars' geometry in the sky. It was done for the tensorial modes without the short wavelength approximation commonly used to simplify the calculations. This work uses that methodology to derive analytical expressions of the overlap reduction functions for the scalar and vectorial polarization modes. In particular, an analytical expression for the scalar-longitudinal mode is determined, which has never been done in prior literature. A comparison of the obtained functions with the latter papers is then provided, which corroborates the detector's sensitivity dependence on the GW frequency for modes with longitudinal components.
\end{abstract}


\maketitle

\section{Introduction}
To clearly identify a signal picked up by pulsar timing arrays (PTA) as originating from a gravitational wave background, the cross correlations must follow a specific pattern, which for the tensor polarizations is known as the Hellings and Downs curve. Modified general relativity can also have vector and scalar polarizations and these would create different such patterns on the cross correlated signals. These patterns are called overlap reduction functions and are calculated as the sky average over all direction dependent terms in the cross correlation. The understanding of the methods for the detection of gravitational wave background from the non-Einsteinian scalar-transverse correlations in PTA data is becoming very relevant. Currently the 12.5-Year data set from the NANOGrav consortium \cite{NANOGrav:2020bcs} found a strong correlation between  various pulsar data. This correlation is consistent across the pulsars with a common stochastic process which is a power-law of same spectral shape and amplitude \cite{NANOGrav:2020bcs}. Later it was found that although the non-Einsteinian scalar-transverse correlations are favoured more than the transverse-traceless correlation but including certain systematics (Solar System ephemeris) and removing just the pulsar J0030+0451 evidence in favour of non-Einsteinian scalar-transverse correlations are dramatically reduced \cite{NANOGrav:2021ini,Chen:2021ncc}.\\

The integration over all sky directions presents some difficulties. The term in the denominator can become zero and thus produces a pole. If the pulsar term is neglected, which is justified in the literature (for example in \cite{Anholm2009,Chamberlin:2011ev,lee}) via short wavelength approximation, then this pole is in general not cancelled. For the purely transverse polarizations the polarization projections vanish as well in the same sky direction, which makes the calculation of this limiting integral possible \cite{lee}. This does not work for polarizations with longitudinal component. As discussed in \cite{lee} the pole is weakened sufficiently by polarization projections in the case of the vector mode such that the integral can be calculated analytically. This is not the case for the longitudinal polarization and only the limiting case for $\phi=0$ could be calculated analytically.\\

We generalize the method we used in our previous work \cite{PTA2} to all possible polarizations in modified GR to calculate the analytic power series of nested sums for the overlap reduction functions for the vector, longitudinal and breathing modes. Since our method does not rely on the short wavelength approximation the poles from the denominator are always sufficiently cancelled by the pulsar term and thus the integral is always well defined. Moreover, our method yields the precise result for all gravitational wave frequencies $\omega$ and all pulsar distances $L$. (This is not required for the current pulsar timing array experiments but nice to have.)\\

Our method thus allows us to calculate the overlap reduction function for the longitudinal polarization analytically for all $\phi$-values for the first time. For the breathing and vector polarizations we find an agreement with Lee et. al. \cite{lee} but disagree with results presented in Chamberlin and Siemens \cite{Chamberlin:2011ev} around the point $\phi=0$.\\
A disadvantage in our result comes from the fact, that it is a power series of nested sums, which is challenging to evaluate for high frequencies. Our generalization in this work allows to extract the main series terms, such that they can be pre-evaluated and then combined with different finite prefactors to form all overlap reduction functions.\\

We review the context in which the overlap reduction functions appear shortly in section (II). In section (III) we present the generalized method from \cite{PTA2} in a summarized way and discuss the results, using plots which depict the most important features in section (IV). Finally, we present our conclusions in section (V).

\section{Overlap reduction functions}\label{sec: ORF}
A general gravitational wave in modified general relativity can be written as:
\begin{align}
    h = \sum_{A}h_A e^A, \qquad A\in\{+,\times,x,y,b,l\},
\end{align}
where $h_A$ is the amplitude and $e^A$ the polarization tensor of polarization $A$.\\

We re-derived the correlated signal for pulsar timing arrays in our previous paper \cite{PTA2}, where we found, that our result agrees with the literature:
\begin{widetext}
\begin{align}
		\mu &= \frac{1}{4\pi^2}\sum_A\int \frac{1}{2\nu^2}S_h^A(\nu)\delta_T^2(\nu-f)
			\frac{1}{4\pi}\int_{\mathbb{S}^2} F^A(\hat{\Omega})F'^A(\hat{\Omega})
			\frac{1-e^{2\pi\mathrm{i}\nu\tau}}{1+\gamma}\frac{1-e^{-2\pi\mathrm{i}\nu\tau'}}{1+\gamma'} d\hat{\Omega}\,
			d\nu\, \tilde{Q}(f) df  \notag \\
		&= \frac{T}{24\pi^2}\int \frac{1}{f^2}\left[\sum_AS_h^A(f)\Gamma_A\right]\tilde{Q}(f) df,
	\end{align}
\end{widetext}
Where we use the pattern functions without poles and discontinuities, which we derived in our first paper on this topic \cite{PTA1}:
\begin{align}\label{eq: FA}
F^+ &= \frac{\alpha^2 - \beta^2}{2}, && F^x = \alpha\gamma, && F^b = \frac{\alpha^2 + \beta^2}{2}, \notag \\
F^\times &= \alpha\beta,  && F^y = \beta\gamma, && F^l = \frac{\gamma^2}{\sqrt{2}},
\end{align}
and collected all the direction dependent terms in the overlap reduction functions:
\begin{align}
	\Gamma_A \coloneq&\, \Lambda\int_{\mathbb{S}} F^{A}(\hat{\Omega})F'^{A}(\hat{\Omega}) \frac{1-e^{\mathrm{i}\frac{L\omega}{c}[1+\gamma]}}{1+\gamma}\frac{1-e^{-\mathrm{i}\frac{L\omega}{c}[1+\gamma']}}{1+\gamma'} d\hat{\Omega},
\end{align}
where $\Lambda=\frac{3}{4\pi}$ is a normalization constant which we use for better comparison with \cite{Chamberlin:2011ev}, where it is called $\beta$.\\

The pattern function $F^A(\hat{\Omega})$ describes the response to GW polarizions of the detector formed with pulsar $a$ which w.l.o.g. lies in direction $\hat{x} = (1,0,0)$ from earth. The response function of pulsar $a'$, which w.l.o.g. lies in direction $\hat{x}' = (\cos\phi,\sin\phi,0)$ is denoted with a prime $F'^A(\hat{\Omega})$.\\

The propagation direction of a gravitational wave introduces a special direction and breaks the rotation symmetry of the problem. We thus end up with an axy-symmetry. A gravitational wave background is assumed to be described by an isotropic power spectral density. The polarization pairs $+$, $\times$ and $x$, $y$ cannot be distinguished from each other in this case, since the axy-symmetry of the problem allows us to change the two polarizations into one another by rotating the basis. Therefore, we form the pairs of tensor $T = \{+,\times\}$ and vector $V = \{x,y\}$ polarizations. This cannot be done with the two scalar polarization, since the rotation around the propagation direction maps $b$ and $l$ into themselves. The longitudinal polarization $l$ is, as the name says, purely longitudinal and the breathing polarization $b$ is purely transverse, which are two different physical phenomenons. Thus we can split the signal accordingly:
\begin{align}
	\mu =& \frac{T}{24\pi^2}\int \frac{1}{f^2}\left[ \sum_{A\in T}S_h^A(f)\Gamma_T \right.\notag \\
	&\left. + \sum_{A\in V}S_h^A(f)\Gamma_V + S_h^b(f)\Gamma_b + S_h^l(f)\Gamma_l \right]\tilde{Q}(f) df,
\end{align}

We calculated the overlap reduction function for the tensor polarizatins $\Gamma_T$ analytically, for general $L\omega$ (i.e. without the assumption that $L\omega<<1$, which is very well justified for the current PTA experiments), by expanding it in a power series. In this work we do the same for the overlap reduction functions of modified general relativity:
\begin{align}
    \Gamma_V \coloneq& \Lambda\int_{\mathbb{S}} 
	\left(F^x(\hat{\Omega})F'^x(\hat{\Omega}) + F^y(\hat{\Omega})F'^y(\hat{\Omega}) \right) \notag \\
	&\qquad \cdot \frac{1-e^{\mathrm{i}\frac{L\omega}{c}[1+\gamma]}}{1+\gamma}\frac{1-e^{-\mathrm{i}\frac{L\omega}{c}[1+\gamma']}}{1+\gamma'} d\hat{\Omega} \\
    \Gamma_b \coloneq& \Lambda\int_{\mathbb{S}} F^{b}(\hat{\Omega})F'^{b}(\hat{\Omega}) \frac{1-e^{\mathrm{i}\frac{L\omega}{c}[1+\gamma]}}{1+\gamma}\frac{1-e^{-\mathrm{i}\frac{L\omega}{c}[1+\gamma']}}{1+\gamma'} d\hat{\Omega}, \\
    \Gamma_l \coloneq& \Lambda\int_{\mathbb{S}} F^{l}(\hat{\Omega})F'^{l}(\hat{\Omega}) \frac{1-e^{\mathrm{i}\frac{L\omega}{c}[1+\gamma]}}{1+\gamma}\frac{1-e^{-\mathrm{i}\frac{L\omega}{c}[1+\gamma']}}{1+\gamma'} d\hat{\Omega},
\end{align}

\section{Power series representation of the overlap reduction functions for scalar and vector polarization's}\label{sec: PowerSeries}
We follow the same methodology as described in our previous work \cite{PTA2} and use the residue theorem, to calculate the integral over the angle $\varphi\in[0,2\pi)$.
\begin{widetext}
	\begin{align}
		\Gamma_M &= \Lambda\sum_{A\in M}\int_0^{2\pi}\int_0^\pi F^A(\hat{\Omega}) F'^A(\hat{\Omega}) \Delta h(\hat{\Omega}) \Delta h'(\hat{\Omega}) \sin\theta d\theta\,d\varphi
			= \Lambda\int_0^\pi \oint_{C_1} f_M(z) dz\,d\theta = 2\pi\mathrm{i}\,\Lambda \int_0^\pi \text{Res}[f_M(z),0] d\theta, \notag \\
		\vphantom{a}\notag\\
		f_M(z) &= \frac{1}{\mathrm{i}z}\sum_{A\in M}\left(F^A {F'}^A\Delta h\Delta h'\right)(\theta,z)\sin\theta,
			\quad \Delta h = \frac{1-e^{\mathrm{i}L\omega[1+\gamma]}}{1+\gamma}, \quad z = e^{\mathrm{i}\varphi},
		\qquad M\in\{V,\{b\},\{l\}\}
	\end{align}
\end{widetext}

After complexifying the integrand we bring it into the same form as described in \cite{PTA2}, eq. (24)-(26):
\begin{equation}
f_M(z) = -\frac{\mathrm{i}e^{-\mathrm{i}\phi}}{8\sin\theta} P_M(z) \left( 1 + E(z) \vphantom{\sqrt{2}}\right)G(z),
\end{equation}
and then split the residue into two parts:
\begin{align}
	\text{Res}[f_M(z),0] =& \underbrace{\text{Res}\left[ -\frac{\mathrm{i}e^{-\mathrm{i}\phi}}{8\sin\theta} P_M(z) G(z) \right]}_{\text{Res}_1} \notag \\
	&+ \underbrace{\text{Res}\left[ -\frac{\mathrm{i}e^{-\mathrm{i}\phi}}{8\sin\theta} P_M(z) EG(z) \right]}_{\text{Res}_2}.
\end{align}

The prefactors of the series coming from the pulsar term contribute to the polynomial term $P_M(z)$ such that it is related to $F^A(z){F'}^A(z)$ by:
\begin{align}
    &P_M(z) = \sum_{N'} A^M_{N'} z^{N'},\\
    &-\frac{\mathrm{i}\,e^{-\mathrm{i}\phi}}{8\sin\theta}P_M(z) = \sum_{A\in M} F^A(z){F'}^A(z)
    \frac{4e^{-\mathrm{i}\phi}z^2}{\sin^2\theta} \frac{\sin\theta}{\mathrm{i}z}
\end{align}
Thus the polarization term is entirely captured by the polynomial term and $G(z)$ and $E(z)$ are the same for every polarization.\\

For the three modes of interest, the polynomial part reads:
\begin{widetext}
    \begin{align}
        %P_l(z) =& \frac{e^{2 \text{i} \phi } \sin ^4\theta }{ z^3}+ \frac{2(1+e^{2\text{i}\phi}) \sin^4\theta}{ z} + e^{-2\text{i}\phi}(1+4e^{2\text{i}\phi}+ e^{4\text{i}\phi})\sin^4\theta\,z + 2(1+e^{-2\text{i}\phi})\,z^3+ e^{-2\text{i}\phi} \sin^4\theta\,z^5 \\ \vphantom{a}\notag\\
        P_l(z) =& \frac{e^{2 \text{i} \phi } \sin ^4\theta }{ z^3}
        + \frac{2(1+e^{2\text{i}\phi}) \sin^4\theta}{ z} + 2(2+\cos(2\phi))\sin^4\theta\,z + 2(1+e^{-2\text{i}\phi})\sin^4\theta\,z^3
        + e^{-2\text{i}\phi} \sin^4\theta\,z^5 \\
        \vphantom{a}\notag\\
        %P_l(z) =& \frac{e^{2 \text{i} \phi } \sin ^4\theta }{ z^3}+ \frac{2(1+e^{2\text{i}\phi}) \sin^4\theta}{ z} + 2(2+\cos(2\phi))\sin^4\theta\,z + 2(1+e^{-2\text{i}\phi})\sin^4\theta\,z^3+ e^{-2\text{i}\phi} \sin^4\theta\,z^5 \\\vphantom{a}\notag\\
        %P_b(z) =& \frac{e^{2 \text{i} \phi } \sin ^4\theta }{2 z^3}+ \frac{(1+e^{2\text{i}\phi})(\cos^4\theta-1)}{z} + \frac{e^{-2\text{i}\phi}}{2}(e^{2\text{i}\phi}(3+\cos(2\theta))^2 + \left( 1 + e^{4\text{i}\phi} \right)\sin^4\theta)\,z \notag \\&+ (1+e^{-2\text{i}\phi})(\cos^4\theta-1)\,z^3+ \frac{1}{2} e^{-2\text{i}\phi} \sin^4\theta\,z^5 \\ \vphantom{a}\notag\\
         P_b(z) =& \frac{e^{2 \text{i} \phi } \sin ^4\theta }{2 z^3}
        + \frac{(1+e^{2\text{i}\phi})(\cos^4\theta-1)}{z}
        + (2(1+\cos^2\theta)^2 + \cos(2\phi)\sin^4\theta)\,z \notag \\
        &+ (1+e^{-2\text{i}\phi})(\cos^4\theta-1)\,z^3
        + \frac{1}{2} e^{-2\text{i}\phi} \sin^4\theta\,z^5 \\
        \vphantom{a}\notag\\
        %P_b(z) =& \frac{e^{2 \text{i} \phi } \sin ^4\theta }{2 z^3} + \frac{(1+e^{2\text{i}\phi})(-2\sin^2\theta+\sin^4\theta)}{z}+ (8-8\sin^2\theta+(2+2\cos(2\phi))\sin^4\theta)\,z \notag \\&+ (1+e^{-2\text{i}\phi})(-2\sin^2\theta+\sin^4\theta)\,z^3+ \frac{1}{2} e^{-2\text{i}\phi} \sin^4\theta\,z^5 \\ \vphantom{a}\notag\\
        %P_V(z) =& -\frac{2 e^{2 \text{i} \phi } \sin ^4\theta }{z^3} + \frac{\left(1+e^{2 \text{i} \phi }\right) \sin ^2(2 \theta )}{z} +2 e^{-2 \text{i} \phi } \left(1+e^{4 \text{i} \phi } + \left(1+4 e^{2 \text{i} \phi }+ e^{4 \text{i} \phi }\right) \cos ^2\theta\right) \sin^2\theta\,z \notag \\ &+ e^{-2 \text{i} \phi } \left(1+e^{2 \text{i} \phi }\right) \sin ^2(2 \theta )\,z^3 - 2 e^{-2 \text{i} \phi } \sin ^4\theta \,z^5\\\vphantom{a}\notag\\
        P_V(z) =& -\frac{2 e^{2 \text{i} \phi } \sin ^4\theta }{z^3}
        + \frac{\left(1+e^{2 \text{i} \phi }\right) \sin ^2(2 \theta )}{z}
        +4 (\cos(2\phi)(1+\cos^2\theta)+2\cos^2\theta) \sin^2\theta\,z \notag \\
        &+ \left(1+e^{-2 \text{i} \phi }\right) \sin ^2(2 \theta )\,z^3
        - 2 e^{-2 \text{i} \phi } \sin ^4\theta \,z^5\\
        % \vphantom{a}\notag\\
      % P_V(z) =& -\frac{2 e^{2 \text{i} \phi } \sin ^4\theta }{z^3} + \frac{4\left(1+e^{2 \text{i} \phi }\right) \sin^2\theta\cos^2\theta}{z} +4 (\cos(2\phi)+(2+\cos(2\phi))\cos^2\theta) \sin^2\theta\,z \notag \\&+ 4\left(1+e^{-2 \text{i} \phi }\right) \sin^2\theta\cos^2\theta\,z^3- 2 e^{-2 \text{i} \phi } \sin ^4\theta \,z^5
    \end{align}
\end{widetext}

As it can be seen, the polynomial part in the three different modes contain the same powers of $z$ as the one for the tensor mode did \cite{PTA2}, (supplemental material) (7). Since $G(z)$ only contains positive powers (see \cite{PTA2} eq. (25)), the first residue is a finite sum:
\begin{align}
    \text{Res}_1^M =& \text{Res}\left[ -\frac{\mathrm{i}e^{-\mathrm{i}\phi}}{8\sin\theta}
    P_M(z)G(z), 0 \right] \notag \\
    =& \text{Res}\left[ \sum_{n\in\mathbb{Z}}b_nz^n, 0 \right] = b_{-1}
\end{align}
For a polynomial starting at $N' = -3$ we get the following residue:
\begin{widetext}
    \begin{align}
        \text{Res}_1^M = \frac{\mathrm{i}e^{-i\phi}}{8\sin\theta}\left( 2e^{-\mathrm{i}\phi}\left[
            \cos\phi - 2\frac{1+2\cos\phi}{\sin^2\theta} \right]A_{-3}^M
            + 2\frac{1+e^{-\mathrm{i}\phi}}{\sin\theta}A_{-2}^M - A_{-1}^M \right)
    \end{align}
\end{widetext}

The contribution of this residue to the overlap reduction function for those three cases then read:\\
\begin{align}
    2\pi\mathrm{i}\int_0^\pi \text{Res}_1^l d\theta =& \frac{2}{3}\pi(3+7\cos\phi) \\
    2\pi\mathrm{i}\int_0^\pi \text{Res}_1^b d\theta =& \frac{\pi}{3}(3+\cos\phi) \\
    2\pi\mathrm{i}\int_0^\pi \text{Res}_1^V d\theta =& -\frac{4}{3}\pi(3+4\cos\phi)
\end{align}

To calculate the second residues we can use the fact, that $\text{Res}$ is a linear map and pull the prefactors out front:
\begin{align}
    \text{Res}_2 =& \text{Res}\left[ -\frac{\mathrm{i}\,e^{-\mathrm{i}\phi}}{8\sin\theta}P_M(z)EG(z),0 \right] \notag \\
    =& -\frac{\mathrm{i}\,e^{-\mathrm{i}\phi}}{8\sin\theta} \sum_{N'} A_{N'}^M S_{N'}, \\
    S_{N'} =& \text{Res}\left[ z^{N'}EG(z),0 \right]
\end{align}
We used this definition of the series terms $S_{N'}$ in the same manner in \cite{PTA2}, (supplemental material), (35). Picking the coefficient to the power $z^{-1}$ leads to the equation:
\begin{align}
    &z^{-n+2j+l+m+2i+N'} = z^{-1} \\
    \Leftrightarrow \quad &n = 2j + l + m + 2i + N, \quad N = N' + 1
\end{align}
where we introduce $N$, which captures the shift in $N'$ due to the power of $-1$ for convenience. We take the explicit expression for the cases $N'>-1$ from \cite{PTA2}, (supplemental material), (34):
\begin{widetext}
	\begin{equation}
    S_{N-1} = \sum_{j=0}^{\infty}\sum_{k=0}^j\sum_{m=0}^k\sum_{l=0}^{j-k}(-1)^je^{-\text{i}(k+m)\phi}\binom{k}{m}\binom{j-k}{l}\left(\frac{2}{\sin\theta}\right)^{j-l-m} \text{i}^{j+l+m+N}g_{j+l+m+N}^{+}(\theta),
    \end{equation}
    and for the special case $N'=-3$ from \cite{PTA2}, (supplemental material), (39):
    \begin{equation}
    \begin{aligned}
     S_{-3} =& \sum_{j=2}^{\infty}\sum_{k=0}^j\sum_{m=0}^k\sum_{l=0}^{j-k}(-1)^je^{-\text{i}(k+m)\phi}\binom{k}{m}\binom{j-k}{l}\left(\frac{2}{\sin\theta}\right)^{j-l-m} \text{i}^{j+l+m-2}g_{j+l+m-2}^{+}(\theta)
     \\&-(1+e^{-2\text{i}\phi})g_0^+(\theta)-\frac{2\text{i}}{\sin\theta}(1+e^{-\text{i}\phi})g_1^-(\theta)-g_2^-(\theta)
     \end{aligned}
     \end{equation}
     where we used the $g_n^\pm(\theta)$ as defined in \cite{PTA2}, (supplemental material), (33):
     \begin{equation}
     \begin{aligned}
    g_n^{\pm}(\theta) &= e^{\text{i}(\text{L}-\text{L}^{\prime})\omega} \left(\frac{\text{L}-e^{\pm\text{i}\phi}\text{L}^{\prime}}{\sqrt{\text{L}^2 + {\text{L}^{\prime}}^2 - 2 \text{LL}^{\prime}\cos\phi}}\right)^n J_n\left(\sqrt{\text{L}^2 + {\text{L}^{\prime}}^2 - 2 \text{LL}^{\prime}\cos\phi}\;\omega\sin\theta\right) 
    \\&- e^{\text{iL}\omega} J_n(\text{L}\omega\sin\theta) - e^{\text{i}(\pm n\phi-\text{L}^{\prime}\omega}) J_n(-\text{L}^{\prime}\omega\sin\theta)
    \end{aligned}
    \end{equation}
	with the Bessel functions $J_n$.
\end{widetext}

We can calculate the contribution of the second residue for all polarizations in an efficient manner, by introducing the functions $H_{N'}^M$:
\begin{align}
    &2\pi\mathrm{i}\int_0^\pi \text{Res}_2\, d\theta = \frac{\pi}{4}e^{-\mathrm{i}\phi} \sum_{N'}H_{N'}^M(\phi), \\
    &H_{N'}^M(\phi) \coloneq \int_0^\pi \sin^{-1}\theta A_{N'}^M S_{N'}\, d\theta
\end{align}
They can be calculated, by expanding the coefficients $A_{N'}^M$ of the polynomial $P_M$ in powers $s$ of $\sin\theta$ and $t$ of $\cos\theta$:
\begin{equation}
    A_{N'}^M = \sum_{s,t} \mathcal{A}_{s,t,N'}^M \sin^{s+1}\theta \cos^t\theta,
\end{equation}
so that we can apply  the identity:
\begin{align}
    &\int_0^{\pi } \cos ^m(\theta ) \sin ^n(\theta ) \, d\theta
    =\frac{\left(1+(-1)^m\right) \Gamma \left(\frac{1+m}{2}\right) \Gamma
       \left(\frac{1+n}{2}\right)}{2 \Gamma \left(\frac{1}{2} (2+m+n)\right)}, \notag \\
    &\text{Re}(m), \text{Re}(n) > -1
\end{align}
in the same way as in our previous work \cite{PTA2}, eq. (28).\\

The coefficients of the extended multi variate polynomials
\begin{equation}
    Q_M = \sum_{N'}\sum_{s,t} \mathcal{A}_{s,t,N'}^M \sin^s\theta \cos^t\theta\,z^{N'}
\end{equation}
in $z$, $\sin\theta$ and $\cos\theta$ for the polarization modes $l$, $b$ and $V$ are given by:\\

\begin{widetext}
\begin{align*}
    &\mathcal{A}_{3,0,-3}^l = e^{2\text{i}\phi}
    &\quad &\mathcal{A}_{3,0,-3}^b = \frac{e^{2\text{i}\phi}}{2} &\quad &\mathcal{A}_{3,0,-3}^V = -2e^{2\text{i}\phi}\\
    &\mathcal{A}_{3,0,-1}^l = 2(1+e^{2\text{i}\phi})
    &\quad &\mathcal{A}_{1,0,-1}^b = -2(1+e^{2\text{i}\phi}) &\quad &\mathcal{A}_{1,2,-1}^V = 4(1+e^{2\text{i}\phi})\\
    &\mathcal{A}_{3,0,1}^l = 2(2+\cos(2\phi))
    &\quad &\mathcal{A}_{3,0,-1}^b = 1+e^{2\text{i}\phi} &\quad &\mathcal{A}_{1,0,1}^V = 4\cos(2\phi)\\
    &\mathcal{A}_{3,0,3}^l = 2(1+e^{-2\text{i}\phi})
    &\quad &\mathcal{A}_{-1,0,1}^b =8 &\quad &\mathcal{A}_{1,2,1}^V = 4(2+\cos(2\phi))\\
    &\mathcal{A}_{3,0,5}^l = e^{-2\text{i}\phi}
    &\quad &\mathcal{A}_{1,0,1}^b =-8 &\quad &\mathcal{A}_{1,2,3}^V = 4(1+e^{-2\text{i}\phi})\\
    & &\quad &\mathcal{A}_{3,0,1}^b =2+2\cos(2\phi)
    &\quad &\mathcal{A}_{3,0,5}^V = -2e^{-2\text{i}\phi}\\
    & &\quad &\mathcal{A}_{1,0,3}^b = -2(1+e^{-2\text{i}\phi}) &\quad\\
    & &\quad &\mathcal{A}_{3,0,3}^b = 1+e^{-2\text{i}\phi} &\quad\\
    & &\quad &\mathcal{A}_{3,0,5}^b = \frac{e^{-2\text{i}\phi}}{2} &\quad
\end{align*}
\end{widetext}

This lead us to the definition in eq. (30) \cite{PTA2}:
\begin{equation}
    h^\pm_{j,b,s,t,N}(\phi) \coloneq \int_0^\pi \sin^{-j+b+s}\theta \cos^t\theta\, g^\pm_{j+b+N}(\theta)\, d\theta
\end{equation}
The special case for which we used $f_{a,b}$ does not occur in the cases of the vector and breathing modes here.\\
%\textcolor{blue}{write about how convenient trigonometric relation can change the coeff such that the condition for integral convergence is always satisfied (and therefore $f_{a,b}$ is not needed) ?}

When we introduce the $\theta$-integrated series term
\begin{widetext}
    \begin{align}
        \mathcal{H}_{J,s,t,N}(\phi) \coloneq \sum_{j=J}^\infty(-2\mathrm{i})^j
        \sum_{k=0}^j\sum_{m=0}^k\sum_{l=0}^{j-k} \left( \frac{\mathrm{i}}{2} \right)^{m+l} \binom{k}{m}
        \binom{j-k}{l} e^{-\mathrm{i}(k+m)\phi} h^+_{j,m+l,s,t,N}(\phi)
    \end{align}
\end{widetext}
we can write the $H_{N'}^M$ explicitly:
\begin{equation}
    H_{N'}^M = \sum_{s,t} \mathcal{A}_{s,t}(\phi) \mathcal{H}_{0,s,t,N}(\phi), \quad N'>-1
\end{equation}
and
\begin{align}
    H_{-3}^M =& \sum_{s,t} \mathcal{A}_{s,t}\left( \mathcal{H}_{2,s,t,N}(\phi)
    + \left[ (1+e^{-2\mathrm{i}\phi})h^+_{0,0,s,t,0}(\phi) \right.\right. \notag \\
    &\left.\left. + 2\mathrm{i}(1+e^{-\mathrm{i}\phi})h^-_{0,0,s-1,t,1}(\phi) + h^-_{0,0,s,t,2}(\phi) \right] \right)
\end{align}
for the special case.\\
Note that instead of collecting all series terms $H_{N'}^M$ in one series as we did last time \cite{PTA2} we leave the series terms seperate. This way we can let the cases $N'>-1$ start from $j=0$ and there is no deed to extract a sum-term to make the series mach with the lower bound of the $H_{-3}^M$-series at $j=2$.\\
As explained in Section IV in our previous paper \cite{PTA2} we reorder the series to increase the efficiency of the numerical evaluation. We transform our indices according to \cite{PTA2}, (supplemental material), eq. (62):
\begin{equation}
    j,k,m,l \ \mapsto \ a = j + b, \ b = m + l, \ k,m
\end{equation}
$ $\\

From \cite{PTA2}, (supplemental material), eq. (63), (65), (68) and (70) we see, that $\mathcal{H}$ transforms as:
\begin{widetext}
    \begin{align}
        \mathcal{H}_{0,s,t,N}(\phi) =& \sum_{a=0}^\infty \sum_{b=0}^{\lfloor \frac{a}{2} \rfloor} \sum_{k=0}^{a-b}
        \sum_{m=\text{max}\{0,2b+k-a\}}^{\text{min}\{k,b\}} (-2\mathrm{i})^{a-b}
        \left( \frac{\mathrm{i}}{2} \right)^b \binom{k}{m} \binom{a-b-k}{b-m} e^{-\mathrm{i}(k+m)\phi}
        h^+_{a,b,s,t,N}(\phi), \\
        \mathcal{H}_{2,s,t,N}(\phi) =& -4e^{-\mathrm{i}\phi}(1+2\cos\phi)h^+_{2,0,s,t,N}(\phi) \notag \\
        &+ \sum_{a=3}^\infty \sum_{b=0}^{\lfloor \frac{a}{2} \rfloor} \sum_{k=0}^{a-b}
        \sum_{m=\text{max}\{0,2b+k-a\}}^{\text{min}\{k,b\}} (-2\mathrm{i})^{a-b}
        \left( \frac{\mathrm{i}}{2} \right)^b \binom{k}{m} \binom{a-b-k}{b-m} e^{-\mathrm{i}(k+m)\phi}
        h^+_{a,b,s,t,N}(\phi)
    \end{align}
\end{widetext}
The exceptional term in $\mathcal{H}_{2,s,t,N}$ for $a=2$ comes from the lower bound of the sum over $a$. When we start from $a = 0$ this term falls away. The $h^+$-functions are the redefined ones from \cite{PTA2}, (supplemental material), eq. (66).\\

Finally the overlap reduction function for the polarization mode $M$ can be written as:
\begin{align}
    \Gamma_M =& 2\pi\mathrm{i}\,\Lambda\left( \int_0^\pi \text{Res}_1\, d\theta + \int_0^\pi \text{Res}_2\, d\theta \right) \notag \\
    =& 2\pi\mathrm{i}\,\Lambda \int_0^\pi \text{Res}_1\, d\theta + \frac{\pi}{4}\Lambda e^{-\mathrm{i}\phi}\sum_{N'}H_{N'}^M,
\end{align}

For the sake of compatibility with \cite{PTA2} we write the expressions for the overlap reduction functions of the vectorial, longitudinal and breathing modes in the same manner as in our previous work i.e. we merge all series-terms into one. We include the normalization constant $\Lambda = \frac{3}{4\pi}$ however:
\begin{widetext}
\begin{equation}
\begin{aligned}
    &\Gamma_l =  \frac{1}{2}(3+7\cos\phi) -\frac{3}{16}e^{\text{i}\phi}\Big[(1+e^{-2\text{i}\phi})h_{0,0,3,0,0}^+(\phi) +2\text{i}(1+e^{-\text{i}\phi})h_{0,0,2,0,1}^-(\phi) + h_{0,0,3,0,2}^-(\phi)\Big]
    \\&+\frac{3}{16}e^{-\text{i}\phi}\pmb{\Bigg[}\sum_{j=0}^1(-2\text{i})^j\sum_{k=0}^j\sum_{m=0}^k\sum_{l=0}^{j-k} \left(\frac{\text{i}}{2}\right)^{m+l}\binom{k}{m}\binom{j-k}{l}e^{-\text{i}(k+m)\phi}
    \\&\cdot\Bigg\{-e^{-2\text{i}\phi}h_{j,m+l,3,0,6}^+(\phi)+2(1+e^{2\text{i}\phi})h_{j,m+l,3,0,0}^+(\phi)+2(1+e^{-2\text{i}\phi})h_{j,m+l,3,0,4}^+(\phi)-2(2+\cos(2\phi))\;h_{j,m+l,3,0,2}^+(\phi)\Bigg\}
    \\&+\sum_{a=2}^{\infty}(-2\text{i})^a\sum_{b=0}^{\lfloor\frac{a}{2}\rfloor}\left(-\frac{1}{4}\right)^{b}
     \\&\cdot\Bigg\{-e^{2\text{i}\phi}h_{a,b,3,0,-2}^+(\phi)-e^{-2\text{i}\phi}h_{a,b,3,0,6}^+(\phi)+2(1+e^{2\text{i}\phi})h_{a,b,3,0,0}^+(\phi)+2(1+e^{-2\text{i}\phi})h_{a,b,3,0,4}^+(\phi)-2(2+\cos(2\phi))\;h_{a,b,3,0,2}^+(\phi)\Bigg\}
    \\&\cdot\sum_{k=0}^{a-b}\begin{cases}
      e^{-\text{i}k\phi}\binom{a-b-k}{b}\; _2F_1(-b,-k;1+a-2b-k;e^{-\text{i}\phi}) & a\geqslant2b+k\\
      e^{\text{i}(a-2(b+k))\phi}\binom{k}{a-2b}\;_2F_1(-a+2b,-a+b+k;1-a+2b+k;e^{-\text{i}\phi}) & a < 2b+k\;\;\;\;\pmb{\Bigg]}
    \end{cases}   
\end{aligned}
\label{eq: GammaLongFinal}
\end{equation}

\begin{equation}
\begin{aligned}
    &\Gamma_b =  \frac{1}{4}(3+\cos\phi) -\frac{3}{32}e^{\text{i}\phi}\Big[(1+e^{-2\text{i}\phi})h_{0,0,3,0,0}^+(\phi) +2\text{i}(1+e^{-\text{i}\phi})h_{0,0,2,0,1}^-(\phi) + h_{0,0,3,0,2}^-(\phi)\Big]
    \\&+\frac{3}{16}e^{-\text{i}\phi}\pmb{\Bigg[}\sum_{j=0}^1(-2\text{i})^j\sum_{k=0}^j\sum_{m=0}^k\sum_{l=0}^{j-k} \left(\frac{\text{i}}{2}\right)^{m+l}\binom{k}{m}\binom{j-k}{l}e^{-\text{i}(k+m)\phi}
    \\&\cdot\Bigg\{-\frac{e^{-2\text{i}\phi}}{2}h_{j,m+l,3,0,6}^+(\phi)+(1+e^{2\text{i}\phi})\left(-2\;h_{j,m+l,1,0,0}^+(\phi) + \;h_{j,m+l,3,0,0}^+(\phi)\right)+(1+e^{-2\text{i}\phi})\left(-2\;h_{j,m+l,1,0,4}^+(\phi)\right.
    \\&\;\;\;\;\;\;\;\;\left.+\;h_{j,m+l,3,0,4}^+(\phi)\right)-8\;h_{j,m+l,-1,0,2}^+(\phi)+8\;h_{j,m+l,1,0,2}^+(\phi)-2(2+\cos(2\phi))\;h_{j,m+l,3,0,2}^+(\phi)\Bigg\}
    \\&+\sum_{a=2}^{\infty}(-2\text{i})^a\sum_{b=0}^{\lfloor\frac{a}{2}\rfloor}\left(-\frac{1}{4}\right)^{b}
    \\&\cdot\Bigg\{-\frac{e^{2\text{i}\phi}}{2}h_{a,b,3,0,-2}^+(\phi)-\frac{e^{-2\text{i}\phi}}{2}h_{a,b,3,0,6}^+(\phi)+(1+e^{2\text{i}\phi})\left(-2\;h_{a,b,1,0,0}^+(\phi)+h_{a,b,3,0,0}^+(\phi)\right)+(1+e^{-2\text{i}\phi})
    \\&\;\;\;\;\;\;\cdot\left(-2\;h_{a,b,1,0,4}^+(\phi) +\;h_{a,b,3,0,4}^+(\phi)\right)-8\;h_{a,b,-1,0,2}^+(\phi)+8\;h_{a,b,1,0,2}^+(\phi)-2(2+\cos(2\phi))\;h_{a,b,3,0,2}^+(\phi)\Bigg\}
    \\&\cdot\sum_{k=0}^{a-b}\begin{cases}
      e^{-\text{i}k\phi}\binom{a-b-k}{b}\; _2F_1(-b,-k;1+a-2b-k;e^{-\text{i}\phi}) & a\geqslant2b+k\\
      e^{\text{i}(a-2(b+k))\phi}\binom{k}{a-2b}\;_2F_1(-a+2b,-a+b+k;1-a+2b+k;e^{-\text{i}\phi}) & a < 2b+k\;\;\;\;\pmb{\Bigg]}
    \end{cases}   
\end{aligned}
\label{GammaBreathFinal}
\end{equation}

\begin{equation}
\begin{aligned}
 &\Gamma_V =  -(3+4\cos\phi) +\frac{3}{8}e^{\text{i}\phi}\Big[(1+e^{-2\text{i}\phi})h_{0,0,3,0,0}^+(\phi) +2\text{i}(1+e^{-\text{i}\phi})h_{0,0,2,0,1}^-(\phi) + h_{0,0,3,0,2}^-(\phi)\Big]
    \\&+\frac{3}{16}e^{-\text{i}\phi}\pmb{\Bigg[}\sum_{j=0}^1(-2\text{i})^j\sum_{k=0}^j\sum_{m=0}^k\sum_{l=0}^{j-k} \left(\frac{\text{i}}{2}\right)^{m+l}\binom{k}{m}\binom{j-k}{l}e^{-\text{i}(k+m)\phi}
     \\&\cdot\Bigg\{2e^{-2\text{i}\phi}h_{j,m+l,3,0,6}^+(\theta)+4\left(1+e^{2\text{i}\phi}\right)h_{j,m+l,1,2,0}^+(\theta)+ 4(1+e^{-2\text{i}\phi})h_{j,m+l,1,2,4}^+(\theta)
    \\&\;\;\;\;\;\;\;\;\;\;\;\;\;\;\;\;\;\;\;\;\;\;\;\;\;\;\;\;\;\;\;\;\;\;\;\;\;\;\;\;-4\cos(2\phi)h_{j,m+l,1,0,2}^+(\theta)-4(\cos(2\phi)+2)h_{j,m+l,1,2,2}^+(\theta)\Bigg\}
    \\&+\sum_{a=2}^{\infty}(-2\text{i})^a\sum_{b=0}^{\lfloor\frac{a}{2}\rfloor}\left(-\frac{1}{4}\right)^{b}
     \\&\cdot\Bigg\{2e^{2\text{i}\phi}h_{a,b,3,0,-2}^+(\theta) +2e^{-2\text{i}\phi}h_{a,b,3,0,6}^+(\theta)+4\left(1+e^{2\text{i}\phi}\right)h_{a,b,1,2,0}^+(\theta)
    \\&+ 4(1+e^{-2\text{i}\phi})h_{a,b,1,2,4}^+(\theta)-4\cos(2\phi)h_{a,b,1,0,2}^+(\theta)-4(\cos(2\phi)+2)h_{a,b,1,2,2}^+(\theta)\Bigg\}
    \\&\cdot\sum_{k=0}^{a-b}\begin{cases}
      e^{-\text{i}k\phi}\binom{a-b-k}{b}\; _2F_1(-b,-k;1+a-2b-k;e^{-\text{i}\phi}) & a\geqslant2b+k\\
      e^{\text{i}(a-2(b+k))\phi}\binom{k}{a-2b}\;_2F_1(-a+2b,-a+b+k;1-a+2b+k;e^{-\text{i}\phi}) & a < 2b+k\;\;\;\;\pmb{\Bigg]}
    \end{cases}    
\end{aligned}
\label{GammaVecFinal}
\end{equation}
where we use the reordered version from \cite{PTA2}, (supplemental material), (66) in the series over $a$:
\begin{equation}
\begin{aligned}
     &h_{a,b,s,t,N}^{\pm}(\phi) \coloneqq \Gamma\left(\frac{t+1}{2}\right)\Gamma\left(b+\frac{s+N+1}{2}\right)
    \\&\cdot\Bigg[e^{\text{i}(\text{L}-\text{L}^{\prime})\omega}\left(\frac{(\text{L}-e^{\pm\text{i}\phi}\text{L}^{\prime})\omega}{2}\right)^{a+N}\;_1\Tilde{F}_2\left(b+\frac{s+N+1}{2};a+N+1,b+\frac{s+t+N}{2}+1;-\left(\text{L}^2 + {\text{L}^{\prime}}^2 - 2 \text{LL}^{\prime}\cos\phi\right)\frac{\omega^2}{4}\right)
    \\&\;\;\;\;\;\;\;\;\;\;\;\;\;\;\;\;\;\;\;\;\;\;\;\;\;\;\;\;\;\;\;\;\;\;\;\;\;\;\;\;\;\;\;\;\;\;\;\;\;\;\;\;\;\;\;-e^{\text{iL}\omega}\left(\frac{\text{L}\omega}{2}\right)^{a+N}\; _1\Tilde{F}_2\left(b+\frac{s+N+1}{2};a+N+1,b+\frac{s+t+N}{2}+1;-\left(\frac{\text{L}\omega}{2}\right)^2\right)
    \\&\;\;\;\;\;\;\;\;\;\;\;\;\;\;\;\;\;\;\;\;\;\;\;\;\;\;\;\;\;\;\;-e^{\text{i}(\pm (a+N)\phi-\text{L}^{\prime}\omega)}\left(-\frac{\text{L}^{\prime}\omega}{2}\right)^{a+N} \;_1\Tilde{F}_2\left(b+\frac{s+N+1}{2};a+N+1,b+\frac{s+t+N}{2}+1;-\left(\frac{\text{L}^{\prime}\omega}{2}\right)^2\right)\Bigg]
\end{aligned}
\end{equation}
\end{widetext}

\section{Discussion of the Results}\label{sec: Discussion}

%For better comparison with \cite{lee} and \cite{Chamberlin:2011ev}, all the overlap reduction functions (eq.\eqref{GammaLongFinal}, eq.\eqref{GammaBreathFinal} and eq. \eqref{GammaVecFinal}) are rescaled with the factor $\beta = \frac{3}{4\pi}$ that was thrown away at the beginning of the calculations.

% First a comparison between the overlap reduction functions truncated series (sums until a suitable cut-off) and direct numerical integrations (using multidimensional rule) is made in Fig.\ref{GammaPhiPlots}, in the case of L = L$^{\prime}$. The computations are made until L$\omega=75$ where it is doable in a reasonable time without requiring a cluster. The truncated series and numerical integrations can be observed to match precisely, thus supporting the correctness of the former. Due to the method used in this work, plotting an analytical curve for the longitudinal mode is possible and new compared to prior literature \cite{lee}\cite{Chamberlin:2011ev}, which only treated the tensorial, vectorial and scalar-breathing modes analytically. As in \cite{PTA2}, the imaginary part of the overlap reduction function for L = L$^{\prime}$ vanishes for all polarization modes. This figure also displays the analytical curves from \cite{lee} for the vectorial and breathing modes, which are the theoretical overlap reduction functions for L $\rightarrow \infty$:

\begin{figure}[h!]
    \begin{minipage}{\linewidth}
        \centering\includegraphics[width=\linewidth]{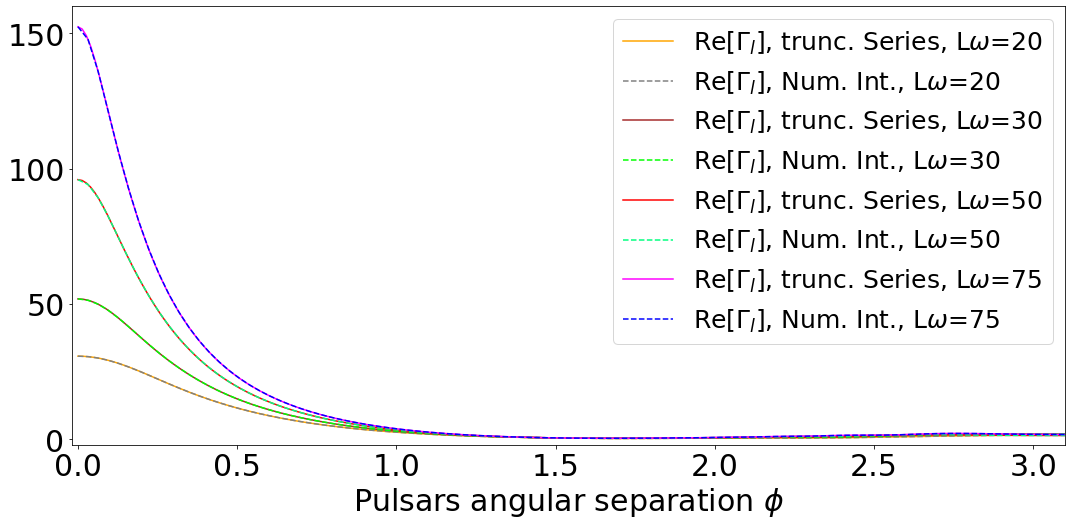}
    \end{minipage}
    \hfill
    \begin{minipage}{\linewidth}
        \centering\includegraphics[width=\linewidth]{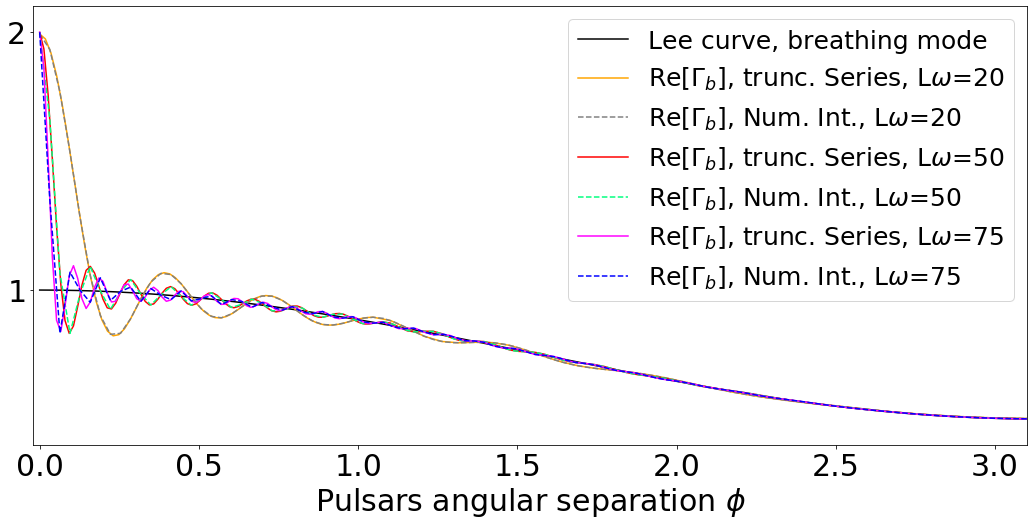}
    \end{minipage}
        \begin{minipage}{\linewidth}
        \centering\includegraphics[width=\linewidth]{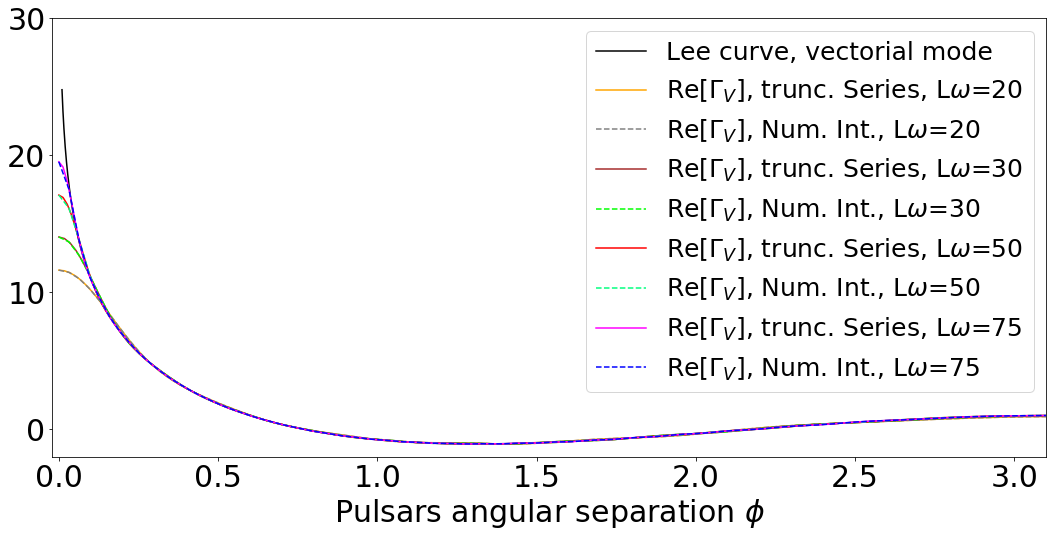}
    \end{minipage}
    \caption{A comparison between the truncated series and numerical integration of the real part of the overlap reduction functions for different $L\omega$ values for the longitudinal (top), breathing (middle) and vector (bottom) polarization. The imaginary parts vanish. For the breathing and vectorial modes, an analytical curve from \cite{lee} is displayed for comparison.}\label{fig: GammaPhiPlots}
\end{figure}

To check the validity of our calculations we compare the truncated series to a direct numerical integration in Fig.~\ref{fig: GammaPhiPlots}. Since the evaluation the series becomes numerically expensive with growing $L\omega$, we plot the $\phi$ dependent overlap reduction functions for a sample of $L\omega$-parameters to visualize its tendency with increasing $L\omega$ up to $75$. We discussed the behaviour of the series for different pulsar distances $L\neq L’$ in detail in our previous paper \cite{PTA2} and focus on the different properties of the overlap reduction functions for the vector, longitudinal and breathing modes in the case of equidistant pulsars $L=L’$ here. In this case the imaginary part vanishes.\\

The limiting overlap reduction functions of the vector and breathing mode for $L\omega\to\infty$ were calculated in \cite{lee}, (A44) and (A33), using the short wavelength approximation. We include the normalization constant $\Lambda$ here for better comparison:

\begin{equation}
    \Gamma_{V,Lee}(\phi) =3\left(-1-\frac{4\cos\phi}{3} + \log\left(\frac{2}{1-\cos\phi}\right)\right)
\label{eq: GammaVLee}
\end{equation}

\begin{equation}
    \Gamma_{b,Lee}(\phi) = \frac{1}{4}(\cos\phi + 3 + 4 \delta(\phi))
\label{eq: GammabLee}
\end{equation}

Since taking the limit of $L\omega\to\infty$ of the power series expression is a daunting if not impossible task, we compare it to the limits obtained by \cite{lee} given in \eqref{eq: GammaVLee} and \eqref{eq: GammabLee} in the middle and lower plot of Fig.~\ref{fig: GammaPhiPlots}.\\
The singularity in the integrand for the longitudinal mode is not sufficiently cancelled by the polarization projections and thus the short wavelength approximation was not applicable. Here we calculate it~\eqref{eq: GammaLongFinal} analytically for the first time and show its graph in the upper plot of Fig.~\ref{fig: GammaPhiPlots}.
\\

%These curves and the truncated series can be seen to match well.

\begin{figure}[h!]
    \begin{minipage}{\linewidth}
        \centering\includegraphics[width=\linewidth]{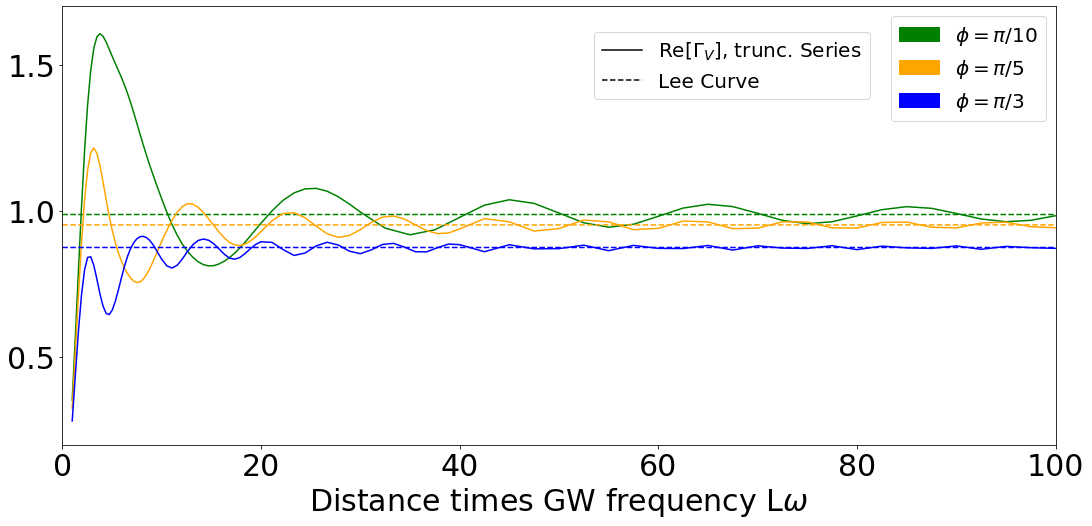}
    \end{minipage}
    \hfill
    \begin{minipage}{\linewidth}
        \centering\includegraphics[width=\linewidth]{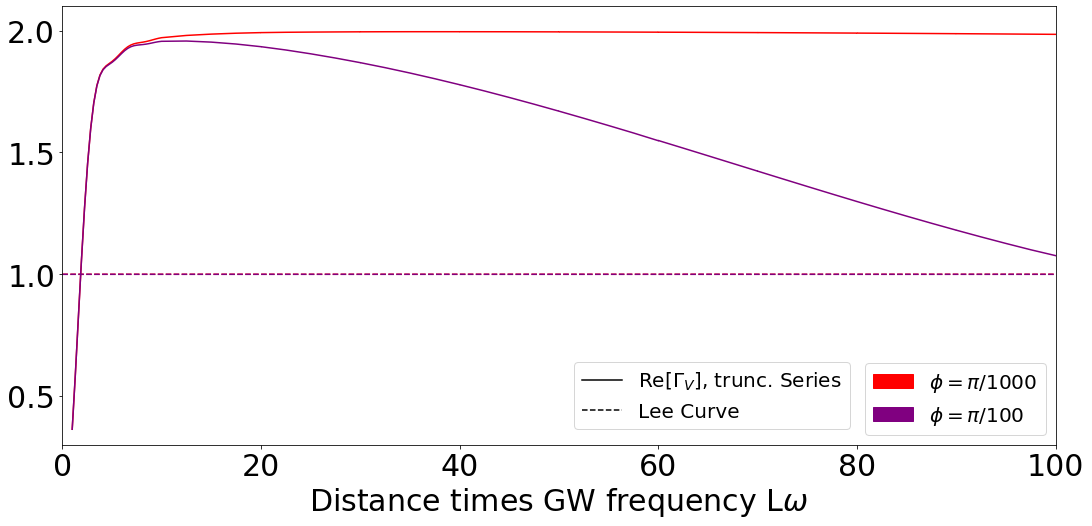}
    \end{minipage}
    \caption{The plot on the top shows the fast convergence of the overlap reduction function for the breathing mode at $\phi$ sufficiently far away from $0$. The bottom plot shows how the function converges slower for $\phi$ close to $0$. The limiting values are calculated from \eqref{eq: GammabLee}.}
    % \caption{Real part of the overlap reduction function truncated series with respect to the pulsars-Earth distance L times the GW frequency $\omega$ for different angular separations $\phi$, in the case of L = L$^{\prime}$, for the breathing mode of the GW. The Lee curve is given by \eqref{eq: GammabLee}. The imaginary part for all those cases vanishes. }
    \label{fig: GammaLoPlotsb}
\end{figure}

%A difference between the polarization mode with purely transverse components (breathing) and the polarization modes that include longitudinal components (longitudinal, vectorial) can be noticed.
As for the tensor polarization mode \cite{PTA2}, the overlap reduction function for the breathing mode converges to $2$ for large $L\omega$ at $\phi=0$, which agrees with the intuition that two aligned detector arms would double the signal.\\
As depicted in Fig.~\ref{fig: GammaLoPlotsb} the function shows the same behaviour like the one for the tensor mode. The upper plot shows, how it converges rapidly to the value obtained using the short wavelength approximation away from $\phi=0$. In the lower plot we show that it converges slower as we approach $\phi=0$ until it assumes the value $2$ at $\phi=0$, in agreement with the $\delta$-term in \cite{lee}.

If the polarization has a longitudinal component, the sensitivity of the detector seems to diverge for co-located pulsars when $L\to\infty$.\\
In appendix A of \cite{PTA1} we found that photons traveling in the same direction as a longitudinal gravitational wave appear to be travelling slightly faster or slower than light speed as seen in the Earth reference frame. This happens, because the gravitational wave is constantly decrease or increase the distance the photons cover by squeezing or stretching the space-time in travel direction. Therefore, two subsequent pulses can increase their time of arrival difference with increasing pulsar difference.\\

%An explanation of such an effect could be found in the appendix of \cite{PTA1} : for 2 photons emitted at the same time from co-located pulsars, the calculations show that from Earth's perspective, they get an effective variation of speed from one another due to the longitudinal part of the gravitational wave. The more distance there is between the pulsars and Earth, the more time the two photons have to separate from one another and the greater their time of arrival difference at the detector becomes. The greater the latter is, the larger the value of the overlap reduction function gets for this configuration, hence explaining its divergence at $\phi=0$ when L $\to\infty$.

\begin{figure}[h!]
    \begin{minipage}{\linewidth}
        \centering\includegraphics[width=\linewidth]{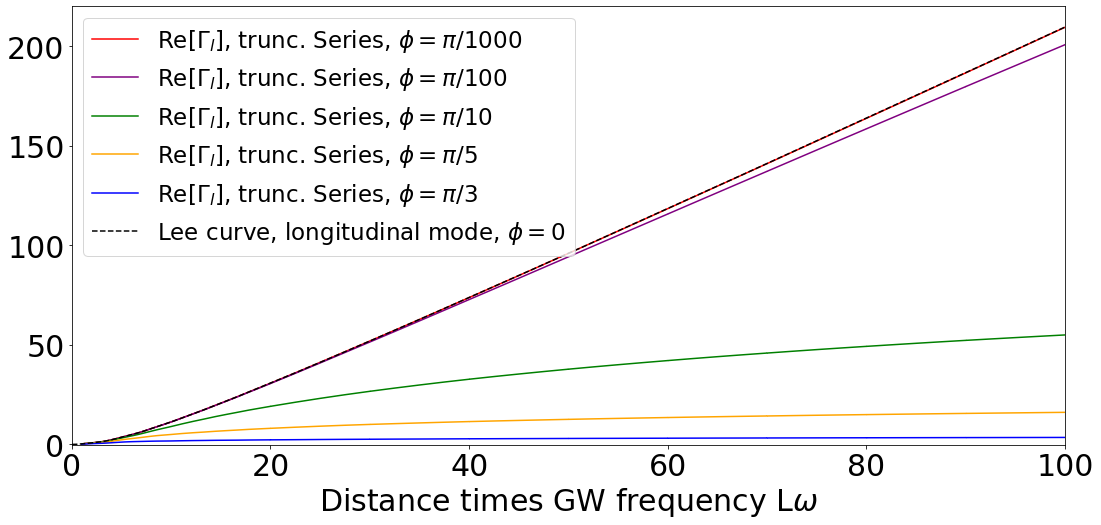}
    \end{minipage}
    \hfill
        \begin{minipage}{\linewidth}
        \centering\includegraphics[width=\linewidth]{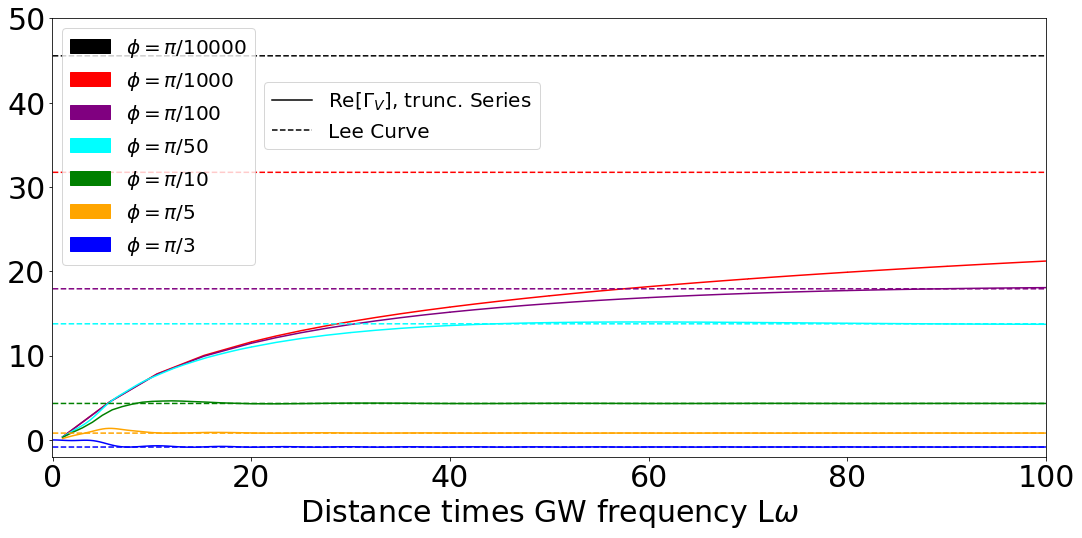}
    \end{minipage}
    \caption{We plot the $L\omega$ dependence of the overlap reduction functions for the longitudinal polarization (top) and the vector polarization (bottom) for $\phi$-values around $0$. We compare our results with the expressions found in \cite{lee} (dashed lines).}
    % \caption{Real part of the overlap reduction function truncated series with respect to the pulsars-Earth distance L times the GW frequency $\omega$ for different angular separations $\phi$, in the case of L = L$^{\prime}$, for the longitudinal and vectorial modes of the GW. For the vectorial case, the Lee curve is given by \eqref{GammaVLee}. The imaginary part for all those cases vanishes. }
    \label{fig: GammaLoPlotslV}
\end{figure}

This feature is more obvious in Fig.\ref{fig: GammaLoPlotslV} and Fig. \ref{fig: GammaLoPlotsb}, where we display the overlap reduction functions dependence on L$\omega$ for different $\phi$ near $0$.
In agreement with \cite{lee,Chamberlin:2011ev} we find that for large $L\omega$ the overlap reduction function for the longitudinal polarization increases linearly for $\phi$-values close enough to $0$.\\
The overlap reduction function for the vector mode converges to increasingly large values with decreasing $\phi$ for $L\omega\to\infty$. Thus, it diverges at $\phi=0$ for $L\omega\to\infty$. This is confirmed by the logarithmic growth in $\Phi\propto L\omega$ shown in \cite{lee}, eq. (A42) at $\phi=0$. In \cite{Chamberlin:2011ev}, Fig. 6 however they claim, that $\Gamma_V$ converges to a finite value with $L\omega\to\infty$ at $\phi=0$.

\section{Conclusions}\label{sec: Conclusions}
We find an analytical expression for the overlap reduction functions of the breathing, longitudinal and vector mode, in form of a power series of nested sums, which is valid for all $L\omega$ since the method does not make any prior assumptions on the gravitational wave frequency or pulsar distance. This is done by generalizing the method we derived in \cite{PTA2}, which does not use the short wavelength approximation which made it possible to calculate the overlap reduction function for the longitudinal polarization for all $\phi$-values for the first time.\\

We formulate the method in a more compact way, which allows us to extract the series terms. This allows to pre-evaluate the series terms and then combine them into the overlap reduction functions for the different modes. This saves computation time since the evaluation of power series of nested sums is numerically costly.\\

Our results are in agreement with prior literature \cite{lee,Chamberlin:2011ev}. We confirm the divergent behaviour of the longitudinal components at $\phi=0$ when $L\omega$ tends to infinity. This should not be interpreted as an infinitely strong signal. Two subsequent pulses acquire a larger time of arrival difference during their journey to earth. Thus, for finite distances the signal will be finite. Also, since the time of arrival increases with $L\omega$ one also requires a longer observation time to detect the two subsequent pulses, which naturally increases the $SNR$ of a persistent signal.\\

The breathing mode behaves similar to the tensor mode and we conclude that the transverse and longitudinal components show a very different behaviour and in the superposition of the two, which is the case for the vector mode the longitudinal feature takes over (since the transverse is finite at $\phi=0$, while the longitudinal grows monotonically with increasing $L\omega$.)\\
The alignment or orthogonality of the polarization with the direction of travel of a wave is a geometrically well-defined (independent of choice of basis) and thus physically relevant property. With that in mind it is not surprising, that for a GWB the two scalar modes, breathing and longitudinal, can be distinguished and the vector $x$, $y$ and tensor $+$, $\times$ cannot.

% This work provided new analytical expansions of the overlap reduction functions for scalar and vectorial modes, using the same methodology as in \cite{PTA2}. This allowed to express those quantities in a purely analytical way without making use of the short wavelength approximation, which strengthens results from prior literature and even gives the first fully analytical expression of the overlap reduction function for the scalar-longitudinal mode. The analysis of those functions lead to similar conclusions to standard literature, namely that for modes with longitudinal components (vectorial $x$, $y$ and scalar $l$), the detector sensitivity is dependent on the GW frequency in the frequency range relevant for PTA. A disagreement with \cite{Chamberlin:2011ev} has nonetheless been found for the vectorial case, where in this work the detector sensitivity has been seen to diverge for co-located pulsars when L$\omega\to\infty$, which is supported by \cite{lee}. For the scalar-breathing mode, the overlap reduction function presented a similar behaviour to the tensorial case derived in \cite{PTA2}, as expected for this purely transverse mode.

% The analytical expressions derived in this work could therefore be used in future GW search with PTA when probing non-Einsteinian metric theories. In particular, they could be useful to study the gravitational wave background, which would give new information about the early universe.

\begin{acknowledgements}
A.B. is supported by the Forschungskredit of the University of Zurich Grant No. FK-21-083 S.T. is supported by Swiss National Science Foundation Grant No. 200020 182047. Symbolic manipulation as well as numerical calculations have been done using Mathematica \cite{Mathematica}.
\end{acknowledgements}

\bibliographystyle{apsrev4-1}
\bibliography{PTARef}

\end{document}